\documentclass{aa}
\usepackage{graphics,epsfig,txfonts,color}
\usepackage[dvips]{rotating}
\usepackage{natbib}
\bibpunct{(}{)}{;}{a}{}{,}

\newcommand{\ergcm}[1]{$\times 10^{#1}$ erg cm$^{-2}$ s$^{-1}$}
\newcommand{\ergcmsr}[1]{$\times 10^{#1}$ erg cm$^{-2}$ s$^{-1}$ sr$^{-1}$}

\newcommand{\ergs}[1]{$\times 10^{#1}$ erg s$^{-1}$}

\newcommand{\hcm}[1]{$\times 10^{#1}$ cm$^{-2}$}

\newcommand{\nh}{N$_{\rm H}$}

\newcommand{\fx}{\hbox{F$_{\rm x}$}}

\newcommand{\HII}{\ion{H}{II}}
\newcommand{\HI}{\ion{H}{I}}

\newcommand{\ltsima}{$\buildrel < \over \sim$}
\newcommand{\lsim}{\lower.5ex\hbox{\ltsima}}
\newcommand{\gtsima}{$\buildrel > \over \sim$}
\newcommand{\gsim}{\lower.5ex\hbox{\gtsima}}

\newcommand{\xmm}{\emph{XMM-Newton}}
\newcommand{\hess}{H.E.S.S.}
\newcommand{\nanten}{\emph{Nanten}}
\newcommand{\chandra}{\emph{Chandra}}
\newcommand{\spitzer}{\emph{Spitzer}}
\newcommand{\fermi}{\emph{Fermi}}
\newcommand{\bepposax}{\emph{BeppoSAX}}
\newcommand{\rosat}{ROSAT}
\newcommand{\hj}{HESS\,J1626$-$490}
\newcommand{\snr}{SNR\,G335.2+00.1}
\newcommand{\rxs}{1RXS\,J162504$-$490918}
\newcommand{\lmxb}{4U\,1624-490}
\newcommand{\hmxb}{IGR\,16283-4838}
\newcommand{\rosatOne}{1RXS\,J162517.7$-$490855}
\newcommand{\OneFGL}{1FGL\,J1626.0$-$4917c}

\begin{document}
\definecolor{orange}{cmyk}{0,.5,1,0}
 
\title{A multi-wavelength study of the unidentified TeV gamma-ray source \hj}
\author{P.\,Eger\inst{1}, 
		G.\,Rowell\inst{2}, 
		A.\,Kawamura\inst{3}, 
		Y.\,Fukui\inst{3}, 
		L.\,Rolland\inst{4}, and
		C.\,Stegmann\inst{1}
	   }

\titlerunning{A multi-wavelength study of \hj}
\authorrunning{Eger et al.}

\institute{\inst{1} Erlangen Centre for Astroparticle Physics, 
				Universit\"at Erlangen-N\"urnberg, Erwin-Rommel-Str. 1, 
				91058 Erlangen, Germany\\
		   \inst{2} School of Chemistry and Physics, University of Adelaide, Adelaide 5005, Australia\\
		   \inst{3} Department of Astrophysics, Nagoya University, 
		   		Furocho, Chikusaku, Nagoya 464-8602, Japan\\
		   \inst{4} CEA Saclay, DSM/IRFU, F-91191 Gif-Sur-Yvette Cedex, France; now at Laboratoire d'Annecy-le-Vieux de Physique des Particules (LAPP), Universit\'e de Savoie, CNRS/IN2P3, F-74941 Annecy-Le-Vieux, France\\
           \email{peter.eger@physik.uni-erlangen.de}
           }
 
%\date{Received dd month 2008 / Accepted dd month 2008}
\date{Received / Accepted }
 
\abstract{} 
         {
		 To explore the nature of the unidentified very-high-energy (VHE, E$>$100~GeV) 
		 gamma-ray source \hj , we investigated the region in X-ray, sub-millimeter, and infrared energy bands. 
		 }
         {
		 So far only detected with the \hess\ array of imaging atmospheric 
		 Cherenkov telescopes, \hj\ could not be unambiguously identified with any source 
		 seen at lower energies. Therefore, we analyzed data from an archival \xmm\ observation, pointed towards \hj , 
		 to classify detected X-ray point sources according to their spectral properties and their 
		 near-infrared counterparts from the 2MASS catalog. 
		 Furthermore, we characterized in detail the diffuse X-ray emission from a region compatible 
		 with the extended VHE signal. 
		 To characterize the interstellar medium surrounding \hj\ we analyzed $^{12}$CO(J=1-0) 
		 molecular line data from the \nanten\ Galactic plane survey, \HI\ data from the Southern 
		 Galactic Plane Survey (SGPS) and \spitzer\ data from the GLIMPSE and MIPSGAL surveys. 
		 }
         {
		 None of the detected X-ray point sources fulfills the energy requirements to be considered as 
		 the synchrotron radiation counterpart to the VHE source assuming an inverse-Compton (IC) emission 
		 scenario. 
		 We did not detect any diffuse X-ray excess emission originating in the region around \hj\ above the 
		 Galactic background and the derived upper limit for the total X-ray flux disfavors a purely leptonic emission 
		 scenario for \hj . 
		 We found a good morphological match between molecular and atomic gas 
		 in the -27km/s to -18km/s line-of-sight velocity range and \hj . 
		 The cloud has a mass of 1.8$\times 10^4$M$_{\odot}$ and is located at a mean kinematic distance of $d$ = 1.8~kpc. 
		 Furthermore, we found a density depression in the \HI\ gas at a similar distance, 
		 which is spatially consistent with the \snr . 
		 We discuss various scenarios for the VHE emission, including the CO molecular cloud being a passive 
		 target for cosmic ray protons accelerated by the nearby \snr . 
		 }
         {}

 
\maketitle

\section{Introduction}
\label{sect-introduction}
During scans of the Galactic plane with the Imaging Atmospheric Cherenkov Telescopes (IACTs) of the 
\hess\ (High Energy Stereoscopic System) array a number of new Galactic very high energy (VHE, E$>$100~GeV) 
$\gamma$-ray sources were detected \citep{2005Sci...307.1938A,2006ApJ...636..777A}. 
Many of these sources could be identified as pulsar wind nebulae (PWN), shell-type supernova remnants (SNR), 
$\gamma$-ray binaries, molecular clouds, or active Galactic nuclei (AGN). 
However, there are still a few TeV sources that could not be unambiguously associated with any source 
detected in lower energy bands \citep[see, e.g.,][]{2008A&A...477..353A}. 
In recent efforts to investigate these `dark' VHE emitters, multi-wavelength observations were conducted 
to probe the environment of these particle accelerators 
\citep[for recent work, see, e.g.,][]{2008A&A...483..509A,2008A&A...481..401A}. 

\hj\ is another VHE gamma-ray source of unknown origin, which so far could not be identified with a source 
at lower wavelengths. 
This object, with an intrinsic extension of $\sim$5~arcmin (Gaussian FWHM), is located right on the Galactic plane 
(R.A.: 16$^h$26$^m$04$^s$, Dec.: -49$^\circ$05\arcmin13\arcsec) and was detected by \hess\ with a peak significance 
of 7.5$\sigma$ \citep{2008A&A...477..353A}. 
It is also stated that the source might be composed of two separate sources owing to the long 
tail extending towards the east. 
These authors measured a power-law spectrum with a photon index of 
2.2~$\pm$~0.1$_{stat}$~$\pm$~0.2$_{sys}$ and a flux normalization of 
\hbox{(4.9$\pm$0.9)$\times$10$^{-12}$cm$^{-2}$\,s$^{-1}$\,TeV$^{-1}$} at 1\,TeV at energies between 0.5~TeV and 40~TeV. 
A number of possible counterparts from other wavelengths are 
discussed by \citep{2008A&A...477..353A}, such as the nearby faint extended \rosat\ source \rxs\ 
\citep{1999A&A...349..389V}, the shell-type \snr\ \citep{1996A&AS..118..329W}, and despite 
their large offsets, the X-ray binaries (XRBs) \lmxb\ and \hmxb . 
Furthermore, the high-energy (HE) source \OneFGL\ from the one-year \fermi\ source catalog 
\citep[1FGL,][]{2010ApJS..188..405A} lies in close proximity to \hj . 

The two physical mechanisms thought to produce VHE gamma rays are inverse-Compton (IC) upscattering 
of low-energy photons by a population of relativistic electrons, as well as production and 
subsequent decay of $\pi ^\circ$s when high-energy hadrons interact with a dense medium. 
The latter, purely hadronic scenario becomes more viable in cases 
where the VHE emission correlates with the location of dense molecular clouds 
\citep[e.g. towards the Galactic Center Ridge,][]{2006Natur.439..695A} or when SNRs interact with 
the surrounding interstellar medium (ISM) as seen with \hess\ from W28 and other SNRs 
\citep{2008A&A...481..401A,2009AIPC.1112...54F} and with VERITAS and MAGIC from IC~443 
\citep{2009ApJ...698L.133A,2007ApJ...664L..87A}. 
In an IC scenario, X-ray emission coinciding with the VHE signal is expected due to synchrotron radiation 
from the same lepton population, whereas only a very low X-ray flux is predicted for a purely hadronic scenario 
arising from synchrotron cooling and/or nonthermal bremsstrahlung from secondary electrons 
produced in the decay of charged pions \citep[see e.g.][]{2009MNRAS.396.1629G}. 

In this work we analyzed the data of an archival \xmm\ observation to search for an X-ray 
counterpart of \hj .
Therefore, we classified the detected X-ray point sources in the vicinity of \hj\ and searched for possible diffuse 
excess emissions above the expected Galactic background. 
Furthermore, we present $^{12}$CO(J=1-0) molecular line survey data taken with the \nanten\ 
mm/sub-mm observatory to scan for molecular clouds and infrared data from the \spitzer\ GLIMPSE and MIPSGAL surveys 
to search for indications of recent star-forming activity.

\section{X-ray data analysis and results}
\label{sect-x-ray}
We retrieved the EPIC (European Photon Imaging camera) data of an archival 
\xmm\ \citep{2001A&A...365L...1J} observation (ID: 0403280201), pointed to the position of \hj . 
The EPIC-MOS \citep{2001A&A...365L..27T} and EPIC-PN \citep{2001A&A...365L..18S} data were analyzed with the 
\xmm\ Science Analysis System (SAS) version 9.0.0, supported by tools from the FTOOLS package and XSPEC 
version 12.5.0 \citep{1996ASPC..101...17A} for spectral modeling. 
For image processing, we used some tools from the CIAO~4.0 software package. 

This observation was affected by long intervals of strong background flaring. 
To clean the data we applied a background good-time-interval (GTI) screening 
based on the full field-of-view (FoV) 7--15~keV light curve provided by the standard processing chain. 
We used thresholds of 8~cts/s for PN and 3~cts/s for MOS and the resulting net exposures 
are 4.9~ks for PN and 13.2~ks for MOS, respectively. 
For all spectra and images presented in the following sections, we selected good (FLAG==0) 
single and multiple events: PATTERN$\leq$4 (PN), PATTERN$\leq$12 (MOS). 

\subsection{X-ray point sources}
\label{sect-point-sources}
\begin{figure}
  \resizebox{0.98\hsize}{!}{\includegraphics[clip=]{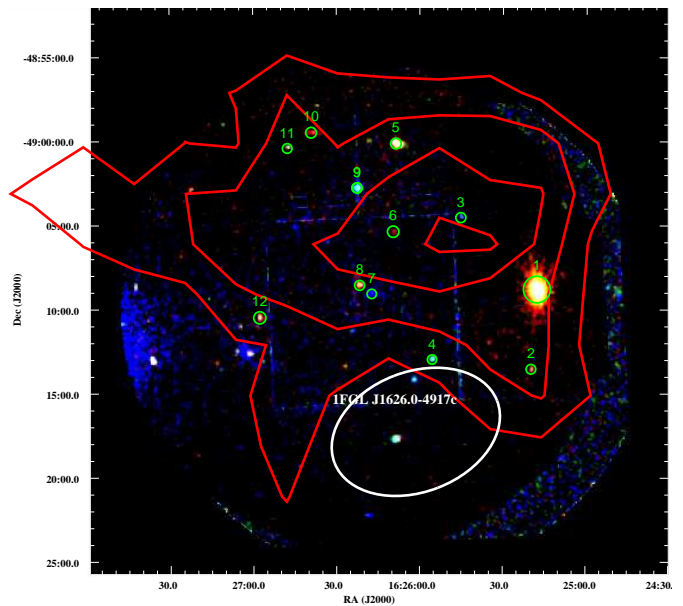}}
  \caption{Combined color-coded image from the PN and MOS detectors. 
  			Energy intervals for the colors are 0.5-1.0\,keV (red), 
			1.0-2.0\,keV (green), and 2.0-4.5\,keV (blue). 
			The red contours denote the 7,6,5,4\,$\sigma$ statistical significance levels of 
			the \hess\ detection of \hj . 
			Marked are point sources detected within the 4$\sigma$~VHE 
			contours (numbered 1 to 12 with increasing R.A.).
			The dashed ellipse (white) shows the 1$\sigma$ confidence region of the positional 
			uncertainty of the \fermi\ source \OneFGL .} 
  \label{fig-xmm-image}
\end{figure}

Even though \citet{2008A&A...477..353A} detected \hj\ with an intrinsic extension of $\sim$0.1$^\circ$, 
we first characterized possible point-like X-ray counterparts. 
The SAS standard maximum likelihood technique for source detection was used in several energy bands: 
0.2--0.5~keV, 0.5--1.0~keV, 1.0--2.0~keV, 2.0--4.5~keV, 4.5--10.0~keV, and 0.5--10.0~keV. 
The resulting point source list indicates a detection threshold of 2\ergcm{-14}. 
Within the 4$\sigma$ contours of \hj , we detected twelve X-ray point sources 
(see Fig.\ref{fig-xmm-image}). 
For five of these sources (1, 5, 7, 8, and 9) the photon statistics were sufficient 
to perform a spectral analysis. 
The remaining seven objects were classified according to their hardness ratio ($HR$), 
which we defined as the normalized difference between high-energy (\emph{Hi}:~2.0--10.0~keV) and low-energy 
(\emph{Low}:~0.2--2.0~keV) counts: 
\begin{equation} HR = \frac{Hi-Low}{Hi+Low}. \label{eq:hr}
\end{equation}
The number of counts in the respective energy bands is the sum over all instruments where 
the source is detected. 
A correction for differential vignetting was applied to the $HRs$. 
Similar to the Galactic plane X-ray source population studies conducted with 
\xmm\ \citep{2004MNRAS.351...31H} and \chandra\ \citep{2005ApJ...635..214E}, we classified 
sources as either soft ($HR<$-0.5), medium (-0.5$<HR<$0.5) or hard ($HR>$0.5). 
To identify near-infrared (NIR) counterparts, we correlated the X-ray point-source positions to the 2MASS all-sky catalog 
\citep{2006AJ....131.1163S}. 
Especially for regions on the Galactic plane where the interstellar absorption is very strong, which is the case for 
\hj\ \citep[\nh$\sim$2.22\hcm{22}][]{1990ARAA...28..215D}, the NIR band is 
particularly suited to probing deep into the interstellar medium (ISM). 
We classified an X-ray source from this observation as NIR-identified when a 2MASS catalog entry is found 
within 2~arcsec of the source position resulting from the SAS source detection algorithm. 
In the following section we classify the detected point sources as either "soft Galactic" (e.g., X-ray active stars), 
"hard Galactic" (e.g., cataclysmic variables (CVs), X-ray binaries), or "hard Extra-galactic" (e.g., background AGN), based on the X-ray spectral characteristics 
and the NIR counterpart. 
These classifications are meant to reflect general characteristics such as hard/soft or absorbed/unabsorbed, which helps for 
deriving more accurate fluxes for sources where no spectral analysis was possible. 

\subsubsection{Soft Galactic sources}
\label{sect-soft}
Owing to coronal activity and/or binary interaction, stars can exhibit strong thermal X-ray emission 
during a wide variety of evolutionary stages 
\citep[for a compilation of the 100 brightest X-ray stars detected by ROSAT, see][]{2003AJ....126.1996M}. 
Therefore, such objects comprise the vast majority of soft Galactic sources. 
A typical physical model to describe their X-ray spectra is an absorbed multi-temperature thermal 
plasma (MEKAL in XSPEC) with temperatures in the range of kT = 0.5--2.0~keV.
Due to the lack of high interstellar absorption at these short distances and their intrinsically soft spectra, these objects would be 
classified as \emph{soft}, according to the $HR$ criterion, mentioned in the previous section. 
Furthermore, we required an NIR counterpart for an identification as a "soft Galactic" source. 
According to these criteria, eight of the twelve point sources were classified as "soft Galactic", 
and a spectral analysis could be performed on three of them. 
We calculated the flux for those sources with no spectrum from the 0.5--10.0~keV count-rate 
assuming an unabsorbed power-law spectrum with photon index 3.0. 
The X-ray and infrared information for these objects are compiled in Table~\ref{tab-classification}. 

\begin{table*}
\begin{center}
\caption[]{Classification of point sources}
\label{tab-classification}
\renewcommand{\arraystretch}{1.3}
\begin{tabular}{llllllllll}
\hline\hline\noalign{\smallskip}
\multicolumn{1}{l}{Source} &
\multicolumn{1}{l}{R.A.} &
\multicolumn{1}{l}{Dec.} &
\multicolumn{1}{l}{\nh} &
\multicolumn{1}{l}{HR$^{(1)}$} &
\multicolumn{1}{l}{kT / $\Gamma ^{(2)}$} &
\multicolumn{1}{l}{F$_{\rm X,unabs}^{(3)}$} &
\multicolumn{1}{l}{J$^{(4)}$} &
\multicolumn{1}{l}{2MASS} &
\multicolumn{1}{l}{class} \\
\multicolumn{1}{l}{No.} &
\multicolumn{2}{l}{} &
\multicolumn{1}{l}{(\hcm{21})} &
\multicolumn{1}{l}{} &
\multicolumn{1}{l}{(for kT: keV)} &
\multicolumn{1}{l}{} &
\multicolumn{1}{l}{(mag)} &
\multicolumn{1}{l}{name} &
\multicolumn{1}{l}{} \\
\noalign{\smallskip}\hline\noalign{\smallskip}
1  	& 16:25:17.5 & -49:08:53 & $<$0.08		 & -0.91 & kT$_1$ = 0.20$^{+0.06}_{-0.02}$ 	&  79.5 	& 6.117  & J16251768-4908524 & soft Galactic\\
	&			 &			 &					&		& kT$_2$ = 0.63$^{+0.02}_{-0.01}$	&			&		 & & \\
	&			 &			 &					&		& kT$_3$ = 2.0$^{+0.09}_{-0.09}$	&			&		 & & \\
2  	& 16:25:19.6 & -49:13:33 & --			& -0.80 & --				   				&  0.39		& 13.458 & J16251980-4913341& soft Galactic\\
3  	& 16:25:45.0 & -49:04:30 & --				& -0.05 & --				   				&  3.5 		& -- 	 & -- & med. extra-galactic \\  
4  	& 16:25:55.4 & -49:12:56 & --				& -0.25 & --				   				&  2.1 		& 12.895 & J16255550-4912576& hard Galactic  \\
5  	& 16:26:08.0 & -49:00:09 & $<$1.6 			& -0.72 & kT$_1$ = 0.46$^{+0.59}_{-0.28}$	&  3.1		& 7.884  & J16260819-4900103& soft Galactic\\
	&			 &			 &					&		& kT$_2$ = 3.5$^{+1.5}_{-1.4}$		&			&		 & -- & \\
6  	& 16:26:09.2 & -49:05:21 & --				& -0.82 & --				   				&  0.044  	& 12.42  & J16260907-4905195& soft Galactic\\
7  	& 16:26:17.2 & -49:09:03 & 88$^{+18}_{-4}$	& +0.92 & $\Gamma$ = 0.41$^{+0.9}_{-0.3}$	&  3.9 		& -- 	 & -- & hard extra-galactic \\
8  	& 16:26:21.6 & -49:08:33 & $<$0.9 			& -0.91 & kT = 0.84$^{+0.3}_{-0.1}$			&  1.9 		& 13.133 & J16262176-4908333& soft Galactic\\
9  	& 16:26:22.3 & -49:02:48 & 6.5$\pm$5.0		& -0.16 & $\Gamma$ = 2.0$\pm$0.7 			&  1.4 		& 11.304 & J16262255-4902485& med. Galactic  \\
10 	& 16:26:38.1 & -48:59:29 & --				& -0.98 & --				   				&  0.091 	& 13.47  & J16263792-4859277& soft Galactic\\
11 	& 16:26:47.2 & -49:00:22 & --				& -0.39 & --				   				&  2.1 		& 12.726 & J16264736-4900230& hard Galactic  \\
12 	& 16:26:57.4 & -49:10:30 & --				& -0.73 & --				   				&  0.35		& 11.704 & J16265766-4910312& soft Galactic\\
\noalign{\smallskip}\hline
\end{tabular}
\end{center}
$^{(1)}$Hardness ratio as defined in the text. 
$^{(2)}$Temperature(s) or photon index resulting from a MEKAL or power-law model fit 
to the spectra; depending on the statistical quality of the thermal spectra, the number 
of temperature components varies. 
$^{(3)}$Unabsorbed flux in the 0.5--10.0~keV band (\ergcm{-13}). The flux is derived from spectral fitting or, 
for fainter sources, by scaling an assumed spectrum with the count-rate depending on the source 
class (see text). 
$^{(4)}$J magnitude from the 2MASS catalog in cases where a counterpart was found within 2~arcsec 
of the X-ray source. 
\end{table*}

\begin{figure}
  \resizebox{0.98\hsize}{!}{\includegraphics[clip=]{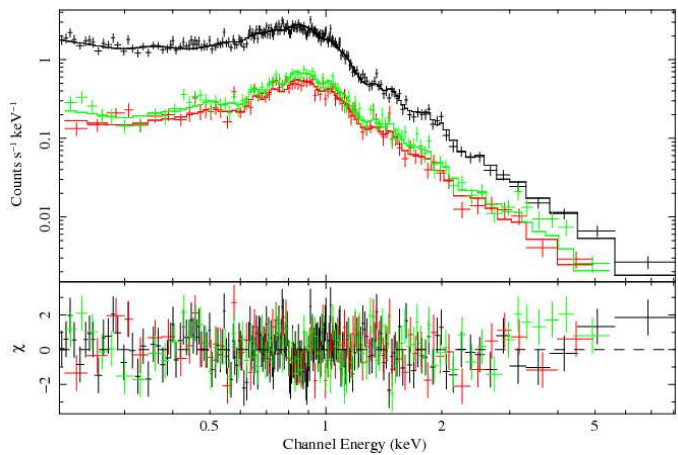}}
  \caption{EPIC-PN and EPIC-MOS spectra of source No.~1 with a 3-temperature thermal plasma model (MEKAL, $\frac{\chi ^2}{\nu}$ = 1.07) fit.}
  \label{fig-xmm-spectrum1}
\end{figure}

Source No.~1 is by far the brightest X-ray point source (\hbox{F$_{\rm X,unabs}$ = 8.0\ergcm{-12}}) 
in the FoV and is spatially coincident with the active triplet system 
\hbox{HD\,147633} = \rosatOne \citep{2003AJ....126.1996M}. 
This system contains a close binary of two G-type main sequence stars. 
The X-ray spectrum (Fig.~\ref{fig-xmm-spectrum1}) can be well-fit with a 
3-temperature thermal plasma emission model (see Table~\ref{tab-classification}). 
The three plasma temperatures resulting from the fit are compatible with temperatures seen, e.g., 
from the Castor X-ray triplet \citep{2001A&A...365L.344G}. 
We therefore conclude that an identification of Source No.~1 with \hbox{HD\,147633} is very likely. 

\subsubsection{Hard Galactic and extra-galactic sources}
\label{sect-hard}
In contrast to the X-ray emission seen from nearby active stars, the spectra of background 
AGN and Galactic CVs are intrinsically harder, and, in the case of AGN, heavily absorbed by 
the interstellar medium in the Galactic plane. 
According to the hardness-ratio classification scheme, these sources would be \emph{medium} or \emph{hard}. 
For two sources, not classified as ``soft Galactic", an X-ray spectrum could be extracted (Nos.~7 and 9). 
Both feature hard power-law spectra. 
Source No.~7 shows a very high absorption column density (\nh\ = 8.8\hcm{22}), 
which is higher than the Galactic level in that direction 
\citep[\hbox{\nh\ = 2.1\hcm{22}},][]{1990ARAA...28..215D}, so we classify this object as ``hard extra-galactic". 
In contrast, the spectrum of the NIR identified source No.~9 is much less absorbed, indicating a location in the 
Galactic plane. 
We therefore classify this source as ``hard Galactic". 
The rest of the \emph{medium} or \emph{hard} sources were classified as ``Galactic" if 
an NIR counterpart is present, otherwise as ``extra-galactic". 
We calculated the flux for sources with no spectrum from the 0.5--10.0~keV count rate 
assuming an absorbed power-law spectrum with photon index 2.0. 
For ``extra-galactic" sources we assumed the total Galactic column density \nh\ = 2.1\hcm{22}, 
for ``Galactic" sources the somewhat lower value \nh\ = 5\hcm{21}. 
The results for all point sources are shown in Table~\ref{tab-classification}. 
We note here that in the rare case of hard Galactic sources with no NIR counterpart, such as PWN, these objects 
would have been wrongly classified as ``extra-galactic".

\subsection{Properties of the diffuse X-ray emission}
\label{sect-diffuse}
\citep{2008A&A...477..353A} approximated the morphology of \hj\ by fitting a 2-D Gaussian to the 
VHE excess map.
The resulting intrinsic (with the effects of the instrument point-spread-function removed) major and minor 
axes were 0.1 deg and 0.07 deg, respectively, with a position angle of 3 deg west to north. 
An IC scenario would predict X-ray emission due to the synchrotron cooling of the same population of electrons. 
Likely candidates for such a scenario are PWN. 
For these sources the X-ray nebula is more compact than the VHE source since the high magnetic 
fields close to the pulsar lead to strong synchrotron cooling. 
In contrast, the VHE nebula might be dominated by a more extended population of electrons with lower energies 
and longer cooling timescales. 
However, as we did not see any obvious extended X-ray counterpart with \xmm\ 
we characterized the diffuse X-ray emission seen from a region comparable to the VHE emission. 

\begin{figure}
  \resizebox{0.98\hsize}{!}{\includegraphics[clip=]{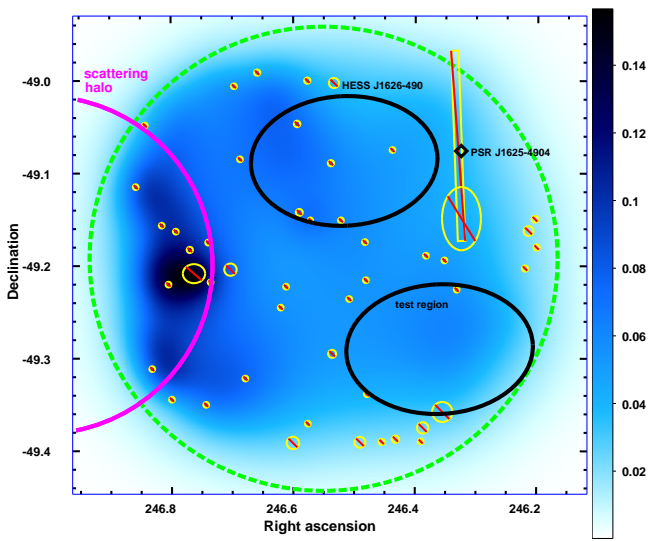}}
  \caption{Adaptively smoothed EPIC-PN image of diffuse X-ray flux in the 
  		   3--7~keV energy band with a linear color scale (arbitrary units). 
		   The instrument FoV is shown as a dashed circle (green). 
		   The large ellipses (black) denote the extraction regions for \hj\ and 
		   the background test, respectively. 
		   The large circle (magenta) to the east gives the approximate extension of the X-ray scattering halo 
		   of \lmxb . 
		   Excluded regions are shown as crossed-out areas (yellow regions crossed with red lines).
		   }
  \label{fig-xmm-diffuse-ima}
\end{figure}

Before the extraction of events from diffuse regions, we first removed all detected point sources from the dataset. 
For that purpose we mostly relied on the results from the SAS maximum-likelihood source detection algorithm. 
Exceptions were very bright sources (e.g. source No.~1), sources at large off-axis angles (e.g. source No.~5), 
and the out-of-time event column from source No.~1, where we modified or added the respective regions manually. 
All exclusion regions are shown in Fig.~\ref{fig-xmm-diffuse-ima}. 

To estimate of the non-X-ray background (NXB) we used a filter-wheel closed dataset, provided by the 
EPIC background working group \citep{2007A&A...464.1155C}. 
To clean the background data of soft proton flares and of contributing observations with a general higher 
flux level, we applied the same GTI threshold as for the source data (see Sect.~\ref{sect-x-ray}). 
However, due to the non-simultaneity of the source and background observations, the particle-induced background level 
could still be different. 
At energies above $\sim$12~keV the effective area of the X-ray telescope decreases significantly and 
the NXB becomes the dominant background component. 
Thus to improve the estimation of the NXB we derived a scaling factor based on the ratio of the 
count-rate in the 12--15~keV energy bands between source and background observations \citep[similar to][]{2004A&A...419..837D}. 

To check for possible hard diffuse X-ray excess emission, we extracted an image in the 3--7~keV energy band.
We refilled excluded regions using \emph{dmfilth} and applied the adaptive smoothing algorithm \emph{csmooth} 
(Fig.~\ref{fig-xmm-diffuse-ima}). 
For the kernel size of the smoothing we required a minimum significance of 5$\sigma$. 
As background we subtracted the scaled NXB dataset as described in the previous paragraph. 

\begin{figure}
  \resizebox{0.98\hsize}{!}{\begin{turn}{-90}\includegraphics[clip=]{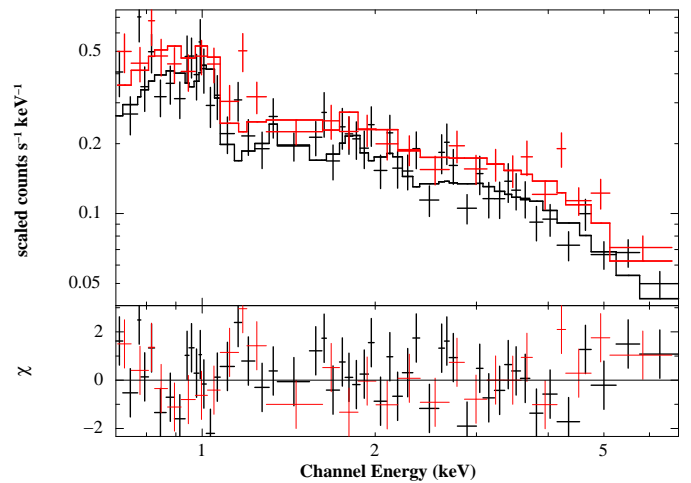}\end{turn}}
  \caption{NXB subtracted spectra from the elliptical region centered at \hj\ (black) and a 
	region in the south of the FoV outside the 4~$\sigma$ contours of \hj\ (red). 
	The stepped lines show the best-fit of a 2-temperature NEI model. 
	}
  \label{fig-xmm-diffuse-spec}
\end{figure}

\subsubsection{Upper limit for extended X-ray emission from \hj}
\label{sect-upper-limit}
The diffuse image (Fig.~\ref{fig-xmm-diffuse-ima}) shows no significant excess emission above the flat Galactic diffuse 
component apart from the area at the eastern edge of the FoV, which is discussed in the next section. 
To derive an upper limit for the total X-ray flux possibly connected to \hj\ (yellow ellipse in Fig.~\ref{fig-xmm-diffuse-ima}) 
we extracted a spectrum from that area and compare it to a spectrum extracted from the southern part of the FoV outside 
the 4~$\sigma$ contours of the VHE detection. 
For the spectral analysis we used the same exclusion regions as for the extraction of the diffuse image (see Sect.~\ref{sect-diffuse}).

To accurately account for the energy-dependent variation of the effective area over these large regions 
we applied the weighting method described by \citet{2001A&A...365L..80A} to extract the spectra. 
In accordance with this method we used the on-axis energy response (RMF) and ancillary response files (ARF). 
As background we subtracted spectra from the identical regions in the NXB dataset using the weighting method as well. 
Both NXB subtracted spectra are shown in Fig.~\ref{fig-xmm-diffuse-spec}. 

Since \hj\ is right on the Galactic plane, we assumed that the observed emission from both regions is of diffuse Galactic origin. 
Therefore, we adopted the two-temperature non-equilibrium ionization model (2-T NEI) from \citet[][henceforth E05]{2005ApJ...635..214E}. 
These authors observed a typical Galactic plane region \citep{1990ARAA...28..215D} with \chandra\ and analyzed the Galactic diffuse emission 
in great detail. 
We fixed most of the model parameters at the values from E05 except for the norm and column densities of both temperature components and 
the Si abundance of the hard component, which would have been significantly overestimated otherwise. 
We fit both spectra in parallel with this model (Fig.~\ref{fig-xmm-diffuse-spec}), and we only allowed the normalizations to vary separately. 
The intrinsic surface fluxes were derived by dividing the unabsorbed fluxes by the effective extraction region, which is the region on the 
detector minus excluded areas, bad pixels/columns, and CCD gaps. 
For both regions the details of the fitting parameters are listed in Table~\ref{tab-twotneifit}. 

\begin{table*}
\caption[]{Fit results for a 2-T NEI model}
\renewcommand{\tabcolsep}{10pt}
\begin{center}
\begin{tabular}{lcc}
\hline\hline\noalign{\smallskip}
\multicolumn{1}{l}{Parameter} &
\multicolumn{1}{l}{\hj} &
\multicolumn{1}{l}{Southern test region} \\
\hline\noalign{\smallskip}
\multicolumn{3}{c}{Soft Component} \\
\noalign{\smallskip}\hline\noalign{\smallskip}
kT (keV)						&	\multicolumn{2}{c}{0.59 (frozen)}		\\ 
log(n$\rm _e$t) (cm$\rm ^{-3}$s)			&	\multicolumn{2}{c}{11.8 (frozen)}		\\ 
Abundance (except Ne, Mg, Si)				&	\multicolumn{2}{c}{0.044 (frozen)}		\\ 
Ne abundance						&	\multicolumn{2}{c}{0.30 (frozen)}		\\ 
Mg abundance 						&	\multicolumn{2}{c}{0.14 (frozen)}		\\ 
Si abundance						&	\multicolumn{2}{c}{0.25 (frozen)}		\\ 
\nh\ (\hcm{22})						&	\multicolumn{2}{c}{0.23$\pm$0.09}		\\ 
Intrinsic surface flux$^{(*)}$ (\ergcmsr{-7})		&	1.4$\pm$0.4 & 1.8$\pm$0.5			\\ 
\hline\noalign{\smallskip}
\multicolumn{3}{c}{Hard Component} \\
\hline\noalign{\smallskip}
kT (keV)						&	\multicolumn{2}{c}{5.0 (frozen)}			\\ 
log(n$\rm _e$t) (cm$\rm ^{-3}$s)			&	\multicolumn{2}{c}{10.6 (frozen)}		\\ 
Abundance (except Fe)					&	\multicolumn{2}{c}{0.17 (frozen)}		\\ 
Fe abundance						&	\multicolumn{2}{c}{0.9 (frozen)}			\\ 
\nh\ (\hcm{22}) 					&	\multicolumn{2}{c}{3.17$^{+0.31}_{-0.27}$} 	\\ 
Intrinsic surface flux$^{(*)}$ (\ergcmsr{-6})			&	2.7$\pm$0.2 & 3.6$\pm$0.3 			\\ 
$\chi ^2$ / $\nu$ 					&	\multicolumn{2}{c}{102 / 74}		\\ 
\noalign{\smallskip}\hline			
\end{tabular}
\end{center}
\begin{center}
$^{(*)}$ In the 0.7--10.0~keV energy band
\end{center}
\label{tab-twotneifit}
\end{table*}

The total surface fluxes for both regions agree within 1.5~$\sigma$, with the flux measured from the 
test region slightly larger. 
Therefore, we did not detect any significant X-ray excess emission from the direction of \hj\ with respect to 
the test region. 
Assuming that a possible X-ray signal associated to \hj\ would originate in a region with the same 
extension as the VHE signal and that an excess flux greater than 4\,$\sigma$ above the Galactic diffuse emission 
would be detectable, we derived an upper limit for the X-ray flux in the 1--10~keV band of 
\hbox{\fx $_{\rm ,excess}$ $<$ 4.85\ergcm{-12}}. 
Given the observed background count rate from the ellipse and assuming an absorbed power-law spectral shape 
with index -2 and column density 1\hcm{22}, the formal 4\,$\sigma$ confidence level flux upper limit 
(1--10~keV) would be $\sim$1\ergcm{-12}, somewhat lower than the above value. 
However, we used the first value as we deem it more realistic for the actual source analysis techniques. 

\begin{figure}
  \resizebox{0.98\hsize}{!}{\begin{turn}{-90}\includegraphics[clip=]{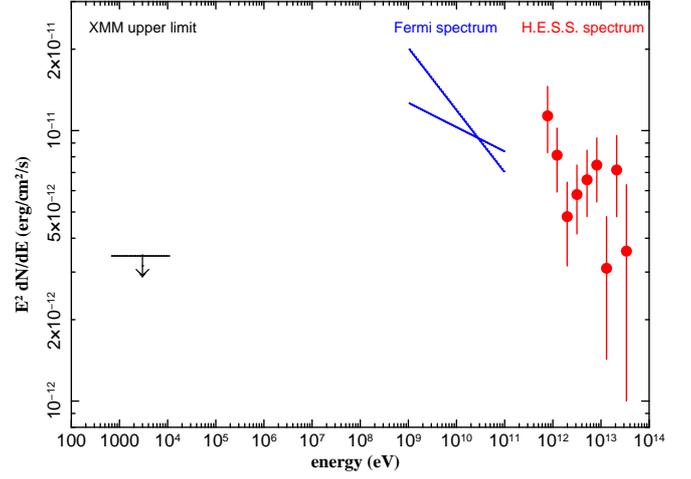}\end{turn}}
  \caption{Spectral energy distribution of \hj , showing the \hess\ spectrum as data points with errors (red) 
		   \citep{2008A&A...477..353A} together with the spectral uncertainty band of \OneFGL\ 
		   \citep{2010ApJS..188..405A}. 
		   The \xmm\ upper limit from this work is denoted by a horizontal line with an arrow (black). 
		   }
  \label{fig-xmm-diffuse-ul}
\end{figure}

Figure~\ref{fig-xmm-diffuse-ul} shows the VHE spectral energy distribution (SED) of \hj , together with 
this upper limit, assuming a power-law spectrum with index -2 for the X-ray spectrum. 
In addition, the spectral uncertainty band of the \fermi\ source \OneFGL\ is shown \citep{2010ApJS..188..405A}.

\subsubsection{Contamination from the dust-scattering halo of the low-mass X-ray binary \lmxb}
\label{sect-scattering-halo}
As seen in the smoothed image of diffuse X-ray emission \ref{fig-xmm-diffuse-ima}, the 
flux level increases towards the eastern edge of the FoV. 
The dipping low-mass X-ray binary \lmxb\ is located $\sim$270~arcsec outside the \xmm\ FoV, exactly 
to the east. 
This source features an extended dust-scattering halo, which was detected with 
\bepposax\ \citep{2000A&A...360..583B} and \chandra\ \citep{2007ApJ...660.1309X}. 
According to these authors the halo shows a hard, highly absorbed spectrum (\nh\ = 8--9\hcm{22}) and should reach 
well inside the \xmm\ FoV of this observation as denoted by the eastern yellow circular region in Fig.~\ref{fig-xmm-diffuse-ima}. 
An absorbed power-law fit to the spectrum extracted from the eastern region 
in the \xmm\ observation gives a column density of \nh\ = (8.0$\pm$2.5)\hcm{22} and a photon index 
of $\Gamma$ = 2.3$\pm$0.7. 
These results, together with the apparent circular shape of the excess, 
confirm the identification of this emission with the dust-scattering halo of \lmxb . 
The effects of this contamination are very limited to the eastern edge of the FoV and should therefore not influence much 
the measurements of the diffuse emission from the direction of \hj\ and the test region. 

\section{\nanten\ $^{12}$CO(J=1-0) data}
\label{sect-nanten}
To search for molecular clouds, spatially and morphologically coincident with \hj , 
we analyzed $^{12}$CO(J=1-0) molecular line observations performed by the 4~m mm/sub-mm 
\nanten\ telescope, located at Las Campanas Observatory, Chile \citep{2004ASPC..317...59M}. 
These data were taken as part of the \nanten\ Galactic Plane Survey \citep[1999 to 2003,][and references therein]{2001PASJ...53.1003M}. 
The survey grid spacing was 4~arcmin for the region around \hj . 
For this work the local standard of rest velocity ($v_{\rm LSR}$) range -240 to +100 km\,s$^{-1}$ was searched. 

\begin{figure*}
 \begin{center}
  \resizebox{0.98\hsize}{!}{\includegraphics[clip=]{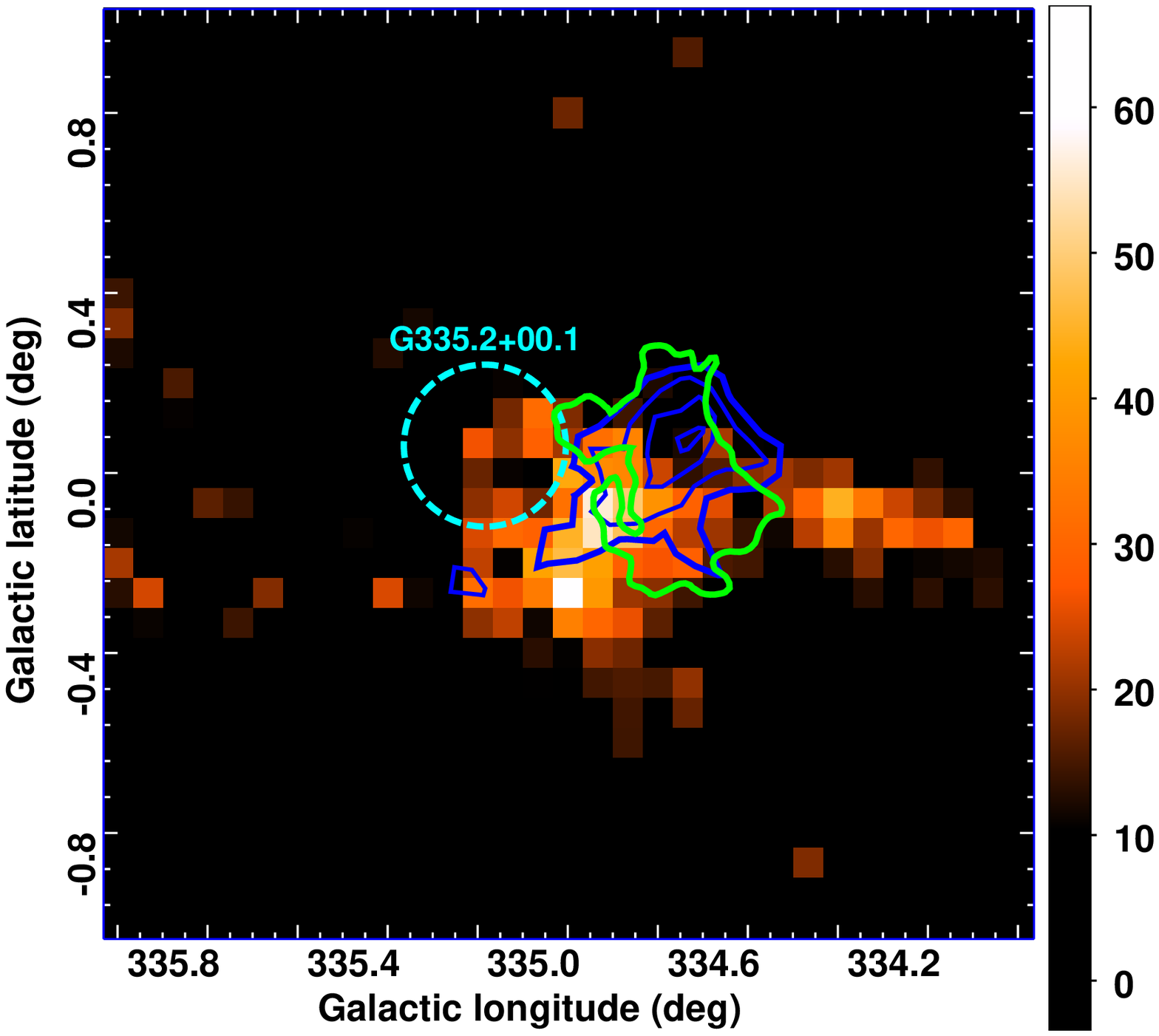}
  		\includegraphics[clip=]{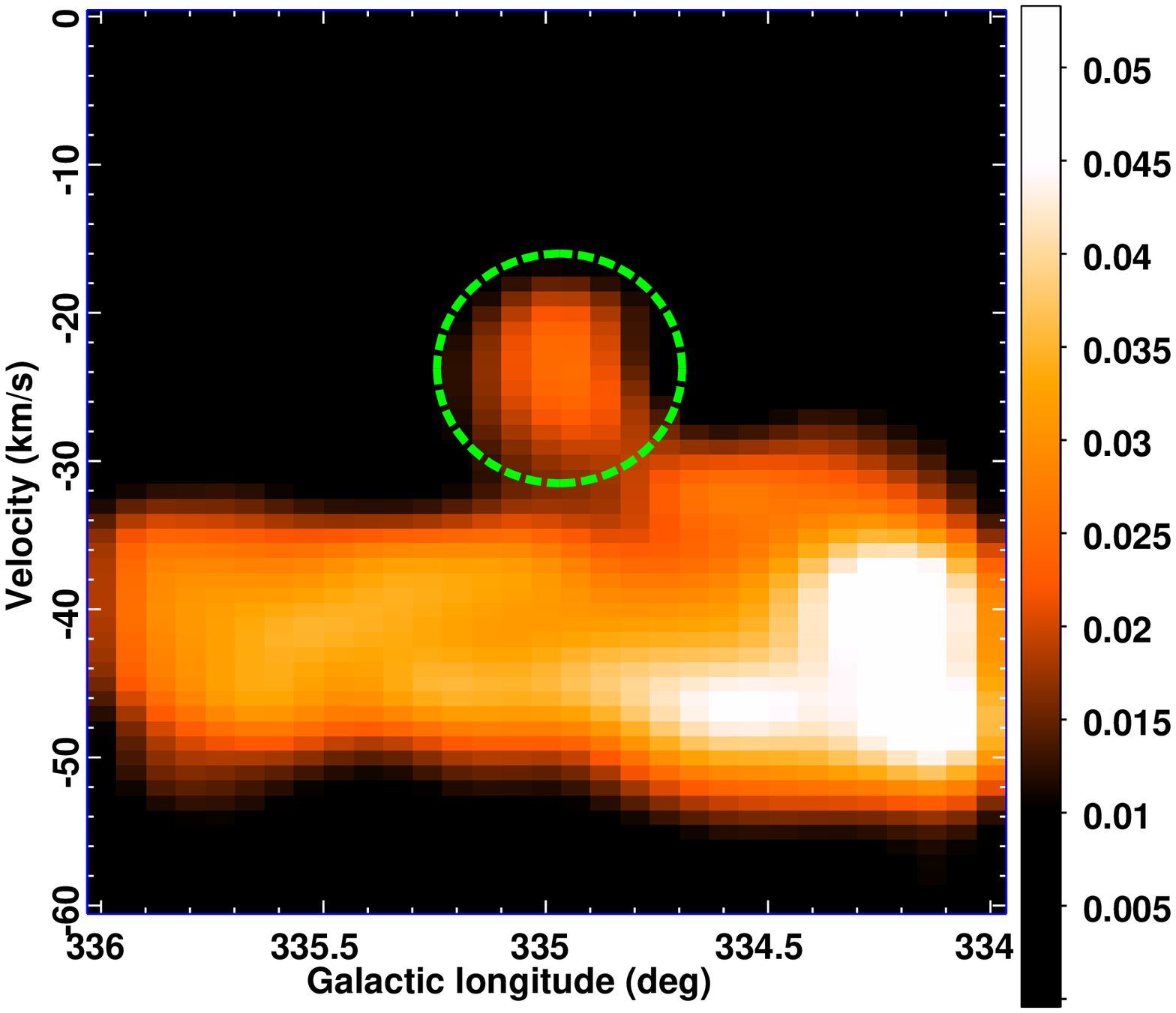}}
  \end{center}
  \caption{\emph{Left:} \nanten\ $^{12}$CO(J=1-0) image of the region around \hj\ (linear scale in K\,km\,s$^{-1}$) 
  		integrated over the $v_{\rm LSR}$ range -31 to -18~km\,s$^{-1}$. 
		Overlaid are the contours of the VHE emission (blue) and of the \HI\ cloud discussed in Sect.~\ref{sect-hiir} 
		(green). 
		The dashed circle (light blue) denotes the position and extension of \snr . 
		\emph{Right:} \nanten\ $^{12}$CO(J=1-0) Galactic longitude--velocity plot (linear scale in K\,deg) 
		integrated over the Galactic latitude range -0.33 to 0.14~deg. 
		The $^{12}$CO cloud shown in the left image is marked by a dashed circle (green). 
  		}
  \label{fig-nanten}
\end{figure*}

Figure~\ref{fig-nanten} (\emph{Left}) shows the \nanten\ $^{12}$CO(J=1-0) image integrated over 
the $v_{\rm LSR}$ range -31 to -18 km\,s$^{-1}$. 
In this interval we found a $^{12}$CO feature partially overlapping with the VHE emission. 
According to the Galactic rotational model of \citet{1993A&A...275...67B}, this $v_{\rm LSR}$ range corresponds 
to a kinematic distance of 2.2 to 1.5~kpc. 
As can be seen in the \nanten\ position-velocity plot (Fig.~\ref{fig-nanten} (\emph{Right})), this CO cloud 
seems to be connected to a much larger system of clouds at velocities -32 to -50~km\,s$^{-1}$. 
The $^{12}$CO emission at these more distant velocities is much more extended, and we did not find any apparent 
feature particularly matching the VHE morphology or coinciding with the \snr .

Using the relation between the hydrogen column density (N$_{\rm H}$) and the $^{12}$CO(J=1-0) 
intensity W($^{12}$CO), \hbox{\nh\ = 1.5 $\times$ 10$^{20}$ [W($^{12}$CO)/K\,km/s]} \citep{2004A&A...422L..47S}, 
we estimate the total mass of this cloud at 1.8$\times 10^4$M$_{\odot}$ for $d$ = 1.8~kpc 
within an elliptical region centered at $l$ = 334.78, $b$ = 0.00 with dimensions 0.26$\times$0.30~deg.
The corresponding average density is 2.1$\times$10$^2$~cm$^{-3}$. 

\begin{figure}
  \resizebox{0.98\hsize}{!}{\includegraphics[clip=]{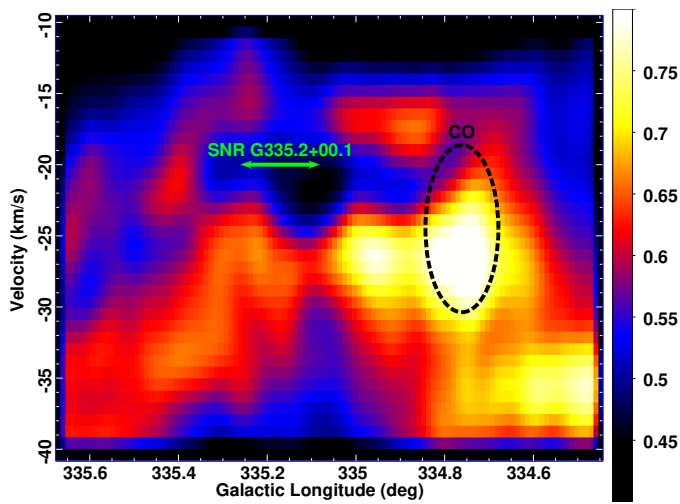}}
  \caption{\HI\ SGPS Galactic longitude--velocity plot (linear scale in K\,deg) 
			integrated over the Galactic latitude range -0.11 to 0.24~deg. 
			The position and extension of the $^{12}$CO cloud is denoted by a dashed circle (black). 
			The double arrow (green) shows the Galactic latitude and extension of \snr . 
  		}
  \label{fig-sgps2}
\end{figure}

\section{HI and infrared data}
\label{sect-hiir}
Figure~\ref{fig-sgps1} shows \HI\ images of the region around \hj\ from the Southern Galactic Plane Survey 
\citep[SGPS][]{2005ApJS..158..178M} integrated over the $v_{\rm LSR}$ ranges -23 to -18~km\,s$^{-1}$ (\emph{Left}) 
and -31 to -23~km\,s$^{-1}$ (\emph{Right}), respectively. 
A local \HI\ density depression is seen in the center of the left image at an angular separation 
of $\sim$21' from the CO molecular cloud and \hj . 
It is striking that this feature is consistent in position as well as in angular extension with the \snr\ 
\citep{2009BASI...37...45G,2006AJ....131.1479R}. 
Furthermore, both images show a region of increased gas density in spatial coincidence with the VHE contours, 
which is more pronounced in the -31 to -23~km\,s$^{-1}$ velocity range. 
Two dips, most likely due to foreground absorption, are marked by arrows (labeled D1 and D2). 

We estimated the mass of the dense \HI\ region coinciding with the VHE and $^{12}$CO features using the signal 
within the same elliptical region as for the CO cloud (Sect.~\ref{sect-nanten}). 
Using the relation between \HI\ intensity and column density from \citet{1990ARA&A..28..215D} 
(X=1.8$\times 10^{18}$\,cm$^{-2}$\,K$^{-1}$\,km/s) we estimated the mass of the cloud as 
4.9$\times 10^{3}$\,${\rm M_{\odot}}$ with an average density of 60.1~cm$^{-3}$. 
However, we note here that due to the two absorption dips in the region this value should be 
seen only as a lower limit of the actual mass of the \HI\ cloud. 

Figure~\ref{fig-sgps2} shows the $v_{\rm LSR}$ profile for this region integrated over the Galactic 
latitude range -0.11 to 0.24~deg, which is the extension of both the \snr\ and the VHE emission of \hj . 
Both, the region of low density and the \HI\ cloud are clearly visible. 
The latter is in good agreement with the position of the CO cloud (Sect.~\ref{sect-nanten}), as indicated 
by the dashed ellipse. 

Figure~\ref{fig-spitzer} shows a three-color (rgb=24/8/3.6 $\mu$m) image from the \spitzer\ GLIMPSE and MIPSGAL 
surveys in log scale (MJy/sr units) with lower limit clipping at (rgb=30/60/5 MJy/sr). 
Apart from the large-scale diffuse emission in the red and green bands along the Galactic plane, there are two 
prominent \HII\ regions within the 4\,$\sigma$ \hess\ contours. 
These two \HII\ regions are coincident with the absorption dips seen in \HI\ (Fig.~\ref{fig-sgps1}), which is a 
common feature of many \HII\ regions in the Galactic plane \citep[see, e.g.,][]{1990ApJ...352..192K}. 
R1 is listed in the catalog of \citet{2003A&A...397..133R} (No.~427) with a systemic distance of 2.4$\pm$0.3~kpc. 
Even though R2 is not listed, judging from the similarity of the two absorption dips, the distances to both 
\HII\ regions should be quite similar. 
Both \HII\ regions are also clearly seen in the Molonglo \citep{1999ApJS..122..207G} 843 MHz radio image 
shown by \citep[][Fig. 2]{2008A&A...477..353A}.
The differing distance estimates for the \HII\ regions and the \HI /CO-complex might indicate 
systematic differences between both methods. 
In particular this could mean that a large fraction of the mass of the \HI\ cloud is located at a distance
more than 1.8~kpc. 

\section{Discussion}
\label{sect-discussion}
In this work we searched for X-ray, sub-millimeter, and infrared counterparts of the unidentified VHE gamma ray source \hj\ to explore 
the nature of the particle acceleration process responsible for the observed emission. 
With a luminosity of 2\ergs{33}$\times$(d/kpc)$^2$, \hj\ is quite typical of Galactic TeV sources detected by \hess , 
and we discuss both leptonic and hadronic emission scenarios in this section.

\begin{figure*}
 \begin{center}
  \resizebox{0.98\hsize}{!}{\begin{turn}{-90}\includegraphics[clip=]{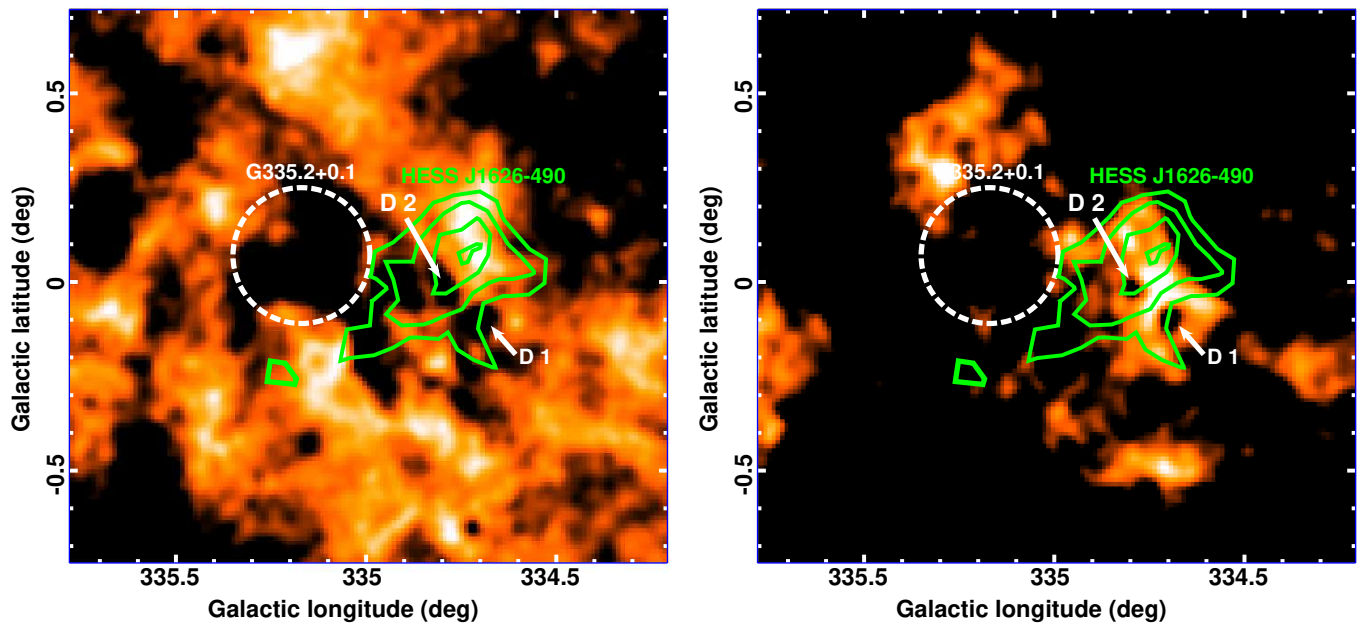}\end{turn}}
  \end{center}
  \caption{\HI\ SGPS images of the region around \hj\ (linear scale in K\,km\,s$^{-1}$) 
  			integrated over the $v_{\rm LSR}$ ranges -23 to -18~km\,s$^{-1}$ (\emph{Left}) 
			and -31 to -23~km\,s$^{-1}$ (\emph{Right}), respectively.
			The dashed circle (white) marks the position and radio extension of \snr . 
			Also shown are the \hess\ VHE contours of \hj\ (green). 
			The arrows (white, labeled D1 and D2) mark two dips, probably due to foreground absorption. 
  		}
  \label{fig-sgps1}
\end{figure*}

According to \citet{2006ApJ...644.1118R}, close binary systems consisting of Wolf-Rayet (WR) and/or OB super-giant stars 
are expected to effectively produce VHE gamma-rays in the colliding wind zone where strong shocks are present and 
where particle acceleration to multi TeV energies could take place. 
The nearby triplet system \hbox{HD\,147633} (source No.~1, Sect.~\ref{sect-soft}) is a strong X-ray emitter. 
Its spectrum can be well fit by a thermal model with temperatures typical of X-ray binaries showing coronal activity. 
This system is composed of a close binary of two G-type main sequence stars and a tertiary star orbiting at a greater 
distance \citep{2003AJ....126.1996M} and can therefore most likely be ruled out as an efficient source of 
highly relativistic particles, which could give rise to VHE gamma-ray emission.
Apart from that, the extended morphology of \hj\ makes an identification with an 
X-ray point-source detected by \xmm\ rather unlikely. 

In the case of a leptonic scenario where low-energy photons are up-scattered by relativistic electrons 
via the IC process, X-ray emission is expected to accompany the VHE signal arising from synchrotron cooling of the 
same population of high-energy electrons \citep{1997MNRAS.291..162A}. 
Two examples for such sources detected by \hess\ are the pulsar wind nebulae MSH\,15-52 
\citep{2005A&A...435L..17A,2010A&A...515A.109S} and HESS\,J1825-13 \citep{2005A&A...442L..25A}. 
Even from a first glance at the flux levels in X-rays ($\sim$\ergcm{-13} for point sources and $\sim$\ergcm{-12} for 
diffuse emission) and at VHE energies ($\sim$\ergcm{-11}), a purely leptonic emission scenario seems unlikely. 
Following \citet{1997MNRAS.291..162A} and assuming a typical Galactic magnetic field strength of 3$\times$10$^{-6}$~G, 
the corresponding synchrotron photon energy for $E$=1~TeV IC photons is $\epsilon$=0.02~keV. 
Assuming the same photon index for the synchrotron counterpart as measured from the VHE source ($\Gamma$ = 2.18) 
and extrapolating the flux to the 0.5 to 10~keV band, we estimate an integrated source flux of $\sim$1.1\ergcm{-11}. 
This flux is a factor of $\sim$25 more than what we measure from any point source, apart from X-ray source No.~1, 
which was discussed in the previous paragraph. 
This makes an identification of \hj\ with any of these sources unlikely. 
Our upper limit for diffuse X-ray emission (Sect.~\ref{sect-upper-limit}) is a factor of $\sim$2 lower 
than the expected value for \hj\ in a purely leptonic model. 

We observed infrared emission from the field around \hj\ (see Sect~\ref{sect-hiir}) 
which could provide an additional target radiation field for the IC process. 
This would lower the expected synchrotron flux in the X-ray band.
On the other hand, these estimates are based on the lowest possible magnetic field found in the Galactic ISM. 
Magnetic fields in the vicinity of energetic pulsars can be a factor of 10--100 larger, which would increase the 
expected synchrotron flux accordingly. 
We therefore conclude that a purely leptonic emission scenario is rather unlikely for \hj . 

Not detecting any X-ray source fulfilling the energetic requirements for a purely leptonic 
scenario favors a hadronic emission process such as dense clouds in the vicinity of a cosmic particle accelerator. 
Such a scenario will be discussed in the remaining part of this section. 
As already mentioned in Sect.~\ref{sect-introduction}, dense molecular clouds are established VHE gamma-ray emitters because 
they provide target material in regions of high cosmic-ray densities. 
Using data from the \nanten\ $^{12}$CO(J=1-0) Galactic plane survey, we detected a 
molecular cloud that is morphologically consistent with \hj . 
This object is located at a kinematic distance of $\sim$1.8~kpc. 
Following \citet[][Eqs. 2 and 3]{1994A&A...285..645A}, we estimated the required gas density to produce 
the observed VHE $\gamma$-ray signal (F$_\gamma$($>$0.6\,TeV) = 7.5$\times$10$^{-12}$~ph~cm$^{-2}$~s$^{-1}$ and 
$\Gamma$ = 2.2) as $n$ $\approx$ 10 cm$^{-3}$ assuming a cosmic ray production efficiency 
of $\theta$ = 0.1 and a distance of d = 1.8 kpc.
This value is an order of magnitude lower than the measured $^{12}$CO mean density. 
Thus, this environment would be easily suited to providing the observed VHE $\gamma$-ray flux. 

\begin{figure}
  \resizebox{0.98\hsize}{!}{\includegraphics[clip=]{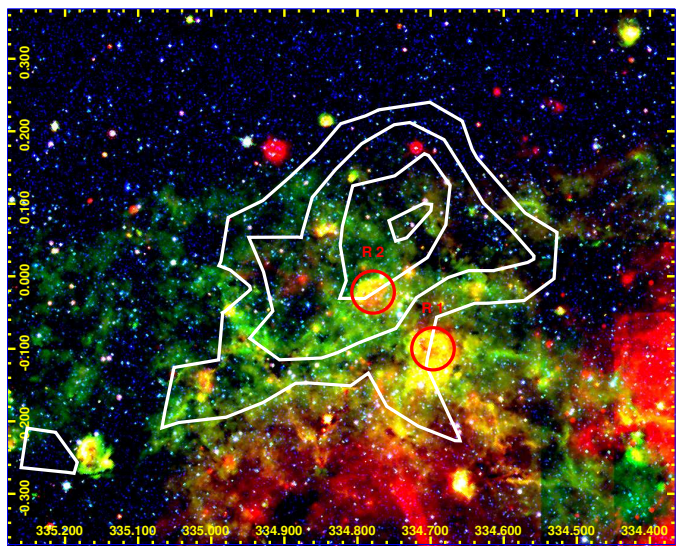}}
  \caption{Three-color (rgb=24/8/3.6 $\mu$m) image from the \emph{Spitzer} 
  		GLIMPSE and MIPSGAL surveys in log scale (MJy/sr units) with
 		lower limit clipping at (rgb=30/60/5 MJy/sr) in Galactic coordinates.
		Shown are the VHE contours of \hj\ (white) and two HII regions (red, 
		labeled R1 and R2). 
		}
  \label{fig-spitzer}
\end{figure}

Following \citet{1991Ap&SS.180..305A} (Eq.~10), assuming that the observed molecular cloud only provides a 
`passive' target for cosmic-rays originating in a nearby acceleration site, the cosmic-ray enhancement factor 
for this source would be $k_{\rm CR}$ = 333 for 10~TeV protons. 
As seen in Fig.~\ref{fig-sgps1} (\emph{Right}) and Fig.~\ref{fig-sgps2} there is a region of 
high-density \HI\ gas spatially coincident with \hj\ located at $v_{\rm LSR}$ $\approx$ -25~km\,s$^{-1}$. 
Taking the additional mass of this \HI\ cloud into account as additional target material 
$k_{\rm CR}$ would be down to $\sim$255 which is an order of magnitude larger than seen from 
other SNR--ISM interactions \citep[see, e.g.,][]{2008A&A...481..401A}. 
However, following \citet{1996A&A...309..917A} (Fig.~1b), even cosmic ray enhancements of $k_{\rm CR}$ $>$ 10$^3$ 
are to be expected from molecular clouds at distances $<$10~pc (as in our case, see next paragraph) from the SNR shell and for 
diffusion coefficients 
of $D$(10\,GeV)$\sim$10$^{26}$~cm$^2$~s$^{-1}$ if the age of the SNR does not exceed 10$^4$ to 10$^5$ years.

In particular in the left image of Fig.~\ref{fig-sgps1}, there appears to be a region of higher concentration of \HI\ gas 
towards the eastern tail of the VHE emission, which in turn could arise from a second VHE source 
(as already mentioned in Sect.~\ref{sect-introduction}) . 
The $^{12}$CO emission that traces the denser regions of the cloud does not show two distinct features 
in a similar way.
As this issue cannot be clearly solved based on the current available VHE data, we did not investigate any subregions 
and only considered the area as a whole.  

Now assuming that this $^{12}$CO / \HI\ cloud is indeed the source of the observed VHE gamma-rays, 
a nearby accelerator would be needed. 
The presence of two prominent \HII\ regions at distances consistent with the CO / \HI\ clouds 
(see Sect.~\ref{sect-hiir}) indicates massive star formation, hence potential PWNe and SNRs in this area. 
The nearby density depression seen in \HI\ (see Sect.~\ref{sect-hiir}) might indicate the presence of a 
recent catastrophic event, such as an SNR, giving rise to strong shocks that would have blown the neutral gas out. 
At d\,=\,1.8~kpc the edge of this region would be at a distance of 8.1~pc from the $^{12}$CO/\HI\ cloud. 
It is striking that this \HI\ feature is consistent in both position and angular extension with the \snr\ 
\citep{2009BASI...37...45G}. 
Using the $\Sigma$-D relation \citet{2004SerAJ.169...65G} estimated the distance to \snr\ as d = 5~kpc.  
which conflicts with the identification of \snr\ with the \HI\ depression seen at a distance of 1.8~kpc.
However, the scatter in $\Sigma$-D distances for individual sources is quite large \citep{2005MmSAI..76..534G} 
making a positive identification of \snr\ and the \HI\ depression seen at a kinematic distance of 1.8~kpc very well possible. 

We note here that, apart from possible systematic uncertainties in kinematic distance estimates, 
the projected distance of 8.1~pc should be seen only as a lower limit because the physical separation 
of accelerator and target could be larger if both objects are not at the same distance to the observer. 
A hint that this could be the case for this source is that the center of gravity of the CO cloud is slightly shifted 
in velocity with respect to the \HI\ density depression. 
A larger distance between the accelerator, most likely \snr , and the molecular cloud would lower the 
expected cosmic ray enhancement.

Even though the angular separation of the \fermi\ source \OneFGL\ and \hj\ is relatively small, these sources are not 
consistent within the 3$\sigma$ significance level of their spatial uncertainties (see Fig.~\ref{fig-xmm-image}). 
Furthermore, we did not find any feature in X-rays, CO, or \HI\ that could explain a displacement between HE and VHE 
gamma-ray emission. 
\OneFGL\ shows signs of a slight variability. 
Based on the variability index listed in the 1FGL catalog \citep{2010ApJS..188..405A}, the chance of the source 
being a constant emitter can be estimated to $<$11\%, which might indicate an extragalactic origin.
Thus it remains unclear at this point whether these two sources are physically connected. 

\section{Outlook}
With the improved angular resolution of next-generation Cherenkov Telescope arrays, it will be possible 
to study the morphology of the VHE signal and its connection to molecular gas in much more detail.
This is crucial for constraining the emission region of VHE $\gamma$-rays, which strongly influences any 
estimation of target mass and density. 
For the determination of diffuse X-ray fluxes and upper limits, a large effective area combined with a good angular 
resolution to effectively exclude point-like sources is needed. 
Therefore, a much deeper exposure ($\sim$100 ks) with \xmm\ would make it possible to significantly improve the constraints 
on leptonic scenarios. 
However, for a more likely hadronic scenario, the flux expectations in the keV band are very low, which could make 
approval of such a proposal very unlikely. 
To study the particle propagation in more detail, dedicated radio polarization observations in the MHz to GHz regime 
would help determine the direction and level of turbulence of the magnetic field in the SNR.
VHE gamma-ray emission from molecular clouds may also exhibit localized peaks at their core regions 
with densities $>$10$^4$~cm$^{-3}$ \citep{2007Ap&SS.309..465G}. 
To trace these regions molecular lines of NH$_3$, CS, or SiO can be used. 
Respective follow-up observations at mm and sub-mm wavelengths could be crucial for characterizing the environment of particle 
interaction.

\begin{acknowledgements}
The XMM-Newton project is supported by the Bundesministerium f\"ur Wirtschaft und 
Technologie/Deutsches Zentrum f\"ur Luft- und Raumfahrt (BMWI/DLR, FKZ 50 OX 0001) 
and the Max-Planck Society. 
This publication makes use of data products from the Two Micron All Sky Survey, which is a joint 
project of the University of Massachusetts and the Infrared Processing and Analysis Center/California 
Institute of Technology, funded by the National Aeronautics and Space Administration and the National 
Science Foundation. 
The NANTEN project is financially supported from JSPS (Japan Society for the Promotion of
Science) Core-to-Core Program, MEXT Grant-in-Aid for Scientific Research on
Priority Areas, and SORST-JST (Solution Oriented Research for Science and
Technology: Japan Science and Technology Agency).
\end{acknowledgements}

\bibliographystyle{aa}
\bibliography{general,myrefereed,mcs,hmxb,ism,ins,cv,hess}

\end{document}